\documentclass[preprint,showpacs,preprintnumbers,amsmath,amssymb]{revtex4}

\usepackage{graphicx}% Include figure files
\usepackage{dcolumn}% Align table columns on decimal point
\usepackage{bm}% bold math
\usepackage{cd}
\usepackage{amsfonts}
\usepackage{amsmath}
\usepackage{times}
\usepackage[T1]{fontenc}
\usepackage[applemac]{inputenc}
\usepackage{fancyhdr}

\newcommand{\px}{\partial_{x}}
\newcommand{\pxb}{\partial_{\bar{x}}}
\newcommand{\pxx}{\partial_{xx}}
\newcommand{\pxxb}{\partial_{\bar{x}\bar{x}}}
\newcommand{\pt}{\partial_{t}}

\begin{document}

\preprint{}

\title{The Burgers equation as electrohydrodynamic model in plasma 
physics}% Force line breaks with \\

\author{\textbf{E. Moreau and O. Vall\'ee}\\
    \emph{\small{ Laboratoire d'Analyse Spectroscopique et
d'\'Energétique des Plasmas\\ Faculté des Sciences, 
rue Gaston Berger  BP 4043 \\
18028 Bourges Cedex France}}}

% \altaffiliation[Also at ]{Physics Department, XYZ University.}
%\affiliation{%

%\affiliation{

\date{\today}% It is always \today, today,

\begin{abstract}
It is shown that the one dimensional Poisson equation 
governing the electric field in a plasma, yields to an inhomogeneous Burgers 
type equation which may be solved analytically for some particular cases. The 
solution obtained for these cases shows an elastic behaviour of the associated 
electric force in the long time limit.
\end{abstract}

%PACS number(s): 47.65.+a, 51.50.+v, 52.30.-q
%\keywords{Burgers equation, electrohydrodynamic, plasma physics}

\maketitle

This paper points out the possibility to treat analytically some 
problems occurring in electro or magnetohydrodynamics (EMHD) models. 
It is well known that the equations describing fluid models are 
difficult to handle because of their mathematical complexities: most of them 
are partial differential equations and present nonlinear properties 
\cite{bec1}, thereby, the analytic approach is nearly totally neglected. 
Nevertheless, there exist some situations where analytic solutions may be 
obtained. One famous example is the Burgers equation 
\cite{burgers1,burgers2}, and especially the homogeneous Burgers 
equation, which is known to give valuable results in turbulence simulations. 
Moreover, this approximation of the Navier-Stokes equation presents some cases 
which are integrable. In particular, the homogeneous Burgers equation in the 
inviscid limit has been studied a lot over past decades, and has given rise to 
the ``Burgulence approach'' \cite{bec2}. But, for a real EMHD problem, 
electric and magnetic fields appear by the way of Maxwell 
equations, so, the problem presents additional difficulties due to the non 
linear coupling between fluid equations and Maxwell equations. A first 
approach in the resolution of such equations has been proposed by some 
authors like Thomas \cite{thomas} and more recently by Olesen 
\cite{olesen2}. In these works, a 1+1 dimensional MHD problem 
is treated with particular configuration for the velocity and the magnetic 
field. The motion equation
\begin{equation}
    \label{b1}
    \pt {\bf{v}}+({\bf{v}}\nabla){\bf{v}}=\nu\nabla^{2}{\bf{v}}+
    (\nabla\times{\bf{B}})\times{\bf{B}},
\end{equation}
is thus coupled with an equation of evolution for the magnetic 
field
\begin{equation}
    \label{b2}
    \pt{\bf{B}}=\nabla\times({\bf{v}}\times{\bf{B}})+
    (1/\sigma)\nabla^{2}{\bf{B}},
\end{equation}
to give a homogeneous Burgers type equation which may consequently 
be integrated. In this model, only the magnetic field is considered, but 
this allows to take into account the Lorentz force acting on a velocity 
field. So, these works are based on a nonlinear connection between an 
equation of motion where appears a magnetic field (Eq. (\ref{b1})), and 
a specific equation for the magnetic field (Eq. ({\ref{b2})). But, another 
interesting aspect of a EMHD problem, consists to study the influence 
of an electric field on a velocity field. Thus, the present paper may be 
considered as a complementary approach facing Olesen's works, in the sense where 
we propose an electric version of Eqs. (\ref{b1}) and (\ref{b2}). In 
ref.\cite{burg} we have solved a Burgers equation with an elastic forcing 
term, which allows to treat a velocity field undergoing a fluctuating strength 
force. Hence, the problem will consist to show that an electric field may be 
considered as an elastic term of force for some physical problems. This will 
be done in the following by using the Poisson equation. In order to 
reach this result, we propose an analytic resolution of a one dimensional 
problem encountered in ionised media, when the electric field is 
dominant in front of the magnetic field. We may have this situation for 
example in electric discharges or electric arcs studies. For this purpose, we 
consider a constant electric field $E_{{\rm{a}}}$ applied to a plasma in which 
we assume the existence of a constant electron flow (created for example by 
electrodes system and depending of $E_{{\rm{a}}}$), which may be interpreted 
as a source term. Moreover, the plasma is assumed to be composed of electrons and 
motionless ions, in such a way that we can write the one dimensional Poisson 
equation as
\begin{equation}
    \label{mhde1}
    \px E=-\frac{\vert e\vert}{\epsilon_{0}}n,
\end{equation}
where $n$ is the electronic density, and $|e|$ the 
absolute value of the electron charge. If the diffusion process is 
taken into account, then the electronic current density will read
\begin{equation}
    \label{mhde2}
    J=-\mu n(E-E_{{\rm{a}}})-\nu\px n;
\end{equation}
with $\mu$ standing for the (constant) electronic mobility, and $\nu$ the 
electronic diffusion coefficient. After what, we use the equation of conservation  
for the electronic density:
\begin{equation}
    \label{mhde3}
    \pt n+\px J=S,
\end{equation}
where $S$ represents the  constant source term. From this, combining  
relations (\ref{mhde1}), (\ref{mhde2}) and (\ref{mhde3}) in order to 
remove the density, we have successively
\begin{equation}
    \label{mhde4}
    J=\frac{\mu\epsilon_{0}}{|e|}(E-E_{{\rm{a}}})\px E+
    \frac{\nu\epsilon_{0}}{\vert e\vert}\pxx E,
\end{equation}
and
\begin{equation}
    \label{mhde5}
    -\frac{\epsilon_{0}}{\vert e\vert}\px\;\pt E+
    \px\left(\frac{\mu\epsilon_{0}}{|e|}(E-E_{{\rm{a}}})\px E+
    \frac{\nu\epsilon_{0}}{\vert e\vert}\pxx 
    E \right)=S;
\end{equation}
which may be integrated to give
\begin{equation}
    \label{mhde6}
    \pt E-\mu(E-E_{{\rm{a}}})\px E=\nu\pxx E-\frac{|e|}{\epsilon_{0}}xS+c(t),
\end{equation}
$c(t)$ being an arbitrary function. After what, carrying out the following 
change of variables 
$$x=\mu \bar{x},\quad
\bar{\nu}=\frac{\nu}{\mu^{2}},\quad  
\bar{S}=S\frac{|e|\mu}{\epsilon_{0}};$$ 
Eq. (\ref{mhde6}) reads
\begin{equation}
    \label{mhde7}
    \pt E-(E-E_{{\rm{a}}})\pxb E=\bar{\nu}\pxxb E-\bar{S}\bar{x}+c(t).
\end{equation}
Thus, the quantity $E_{{\rm{a}}}$ being constant, a 
Burgers type equation may be obtained for the variable 
$E-E_{{\rm{a}}}$:
\begin{equation}
    \label{mhde8}
    \pt (E-E_{{\rm{a}}})-(E-E_{{\rm{a}}})\pxb (E-E_{{\rm{a}}})=
    \bar{\nu}\pxxb(E-E_{{\rm{a}}})-\bar{S}\bar{x}+c(t).
\end{equation}
The local field $E(\bar{x},t)$ may then be determined by putting 
\begin{equation}
    \label{mhde9}
    \mathcal{E}(\bar{x},t)=E(\bar{x},\bar{t})-E_{{\rm{a}}}, 
\end{equation}    
that is to say that we have to solve the following one dimensional Burgers 
equation with an elastic forcing term 
\begin{equation}
    \label{mhde15}
    \pt \mathcal{E}-\mathcal{E}\pxb \mathcal{E}=
    \bar{\nu}\pxxb \mathcal{E}-\bar{x}\bar{S}+c(t);
\end{equation}
with 
\begin{equation}
    \label{mhde5}
    \mathcal{E}(\bar{x},0)=-E_{{\rm{a}}}
\end{equation}
as initial condition if initially there is no charges separation. 
Notice that a minus sign appears with the non linear term, with the 
meaning that the electrons spread in the opposite direction of the electric 
field. Then, We have shown ref.\cite{burg} the possibility to solve analytically 
such an equation by two equivalent methods. Putting $c(t)$ to zero, it 
is hence possible to show that the solution reads, in physical coordinates $(x,t)$,
\begin{equation}
    \label{sole4}
    \mathcal{E}(x,t)=-\frac{\gamma}{\mu}
    \left(1-e^{-2\gamma t}\right)x- E_{{\rm{a}}}e^{-\gamma t}
\end{equation}
where we have put
\begin{equation}
    \gamma=\sqrt{\frac{S|e|\mu}{\epsilon_{0}}}.
\end{equation}
It must be underlined that the effective electric field acting 
on any electron of the system is in fact $-\mathcal{E}(x,t)$. 
With this in mind, a physical interpretation of the problem may be 
obtained in the asymptotic mode when $S\ge 0$ ({\it{i.e}} 
$\gamma\in\mathbb{R}^{+})$. In this case, the asymptotic limit, reached for 
$t\gg 1/\gamma$, simply reads
\begin{equation}
    \label{sole7}
   \lim_{t\to\infty}\mathcal{E}(x,t)=-\frac{\gamma}{\mu}\;x,
\end{equation}
which is a stationary solution. In other words, the system has 
reached a stationary state. Thus, we  can deduce that the electric force $F$ 
acting on an electron is related to an elastic 
forcing term. As a matter of fact, we have in this stationary state
\begin{equation}
    \label{sole8}
    F=-|e|\times-\lim_{t\to\infty}\mathcal{E}(x,t)=
    -|e|\frac{\gamma}{\mu}x=-\kappa^{2} x,
\end{equation}
where $\kappa^{2}$ stands for a string constant and is given by
\begin{equation}
    \label{sole9}
    \kappa^{2}=|e|\frac{\gamma}{\mu}=
    \left(\frac{S|e|^{3}}{\mu\epsilon_{0}}\right)^{1/2}.
\end{equation}
Notice that an electron undergoing the single force (\ref{sole8}) 
(low viscosity), will oscillate with a pulsation $\omega$ such as
\begin{equation}
    \omega^{2}=\frac{\kappa^{2}}{m_{e}}=
   \frac{|e|\gamma}{\mu m_{e}};
\end{equation}
$m_{e}$ standing for the electronic mass. This expression has to be 
compared with the plasma frequency $\omega_{{\rm{p}}}$:
\begin{equation}
    \label{frec1}
    \omega^{2}_{{\rm{p}}}=\frac{|e|^{2}n}{\epsilon_{0}m_{e}},
\end{equation}
with $n$ given by the Poisson equation (\ref{mhde1}). In the long time 
limit we have
\begin{equation}
    \label{frec2}
    n=-\frac{\epsilon_{0}}{|e|}\px\lim_{t\to\infty}E(x,t)=
    \frac{\epsilon_{0}\gamma}{|e|\mu}.
\end{equation}
So,
\begin{equation}
    \label{frec3}
    \omega^{2}_{{\rm{p}}}=\frac{|e|\gamma}{\mu m_{e}}=\omega^{2}.
\end{equation}
Hence, for a low viscosity system, the electrons of our model oscillate with the 
theorical plasma frequency. In a more general case, when 
viscosity effects have to be taken into consideration, we assume the following 
motion equation statisfied for the electronic velocity $u$
\begin{equation}
    \label{burg1}
    \pt u+u\px u=\nu\pxx u-\kappa^{2} x.
\end{equation}
In this case, an electron undergoes both viscosity and electric 
field. The Eq. (\ref{burg1}) is an inhomogeneous Burgers equation with an 
elastic term of force. But, we have shown ref.\cite{burg} that in such 
a case, a physical solution was
\begin{equation}
    \label{burgg}
    \lim_{t\to\infty}u(x,t)=\kappa x.
\end{equation}
Then, assuming that the velocity $u$ writes (relation (\ref{mhde2}))
\begin{equation}
    \label{u1}
    u(x,t)=-\mu\mathcal{E}(x,t)-\nu\px\ln[n(x,t)],
\end{equation}
with $n(x,t)$ given by the Poisson equation (\ref{mhde1}), we have in the 
asymptotic limit
\begin{equation}
    \lim_{t\to\infty} u(x,t)=-\mu\lim_{t\to\infty}\mathcal{E}(x,t)-
    \nu\lim_{t\to\infty}\px\ln[n(x,t)].
\end{equation}
Then, a relation connecting the source term with the electronic mobility 
appears from the limits (\ref{sole7}) and (\ref{burgg}):
\begin{equation} 
    \label{u3}
    S\mu^{3}=|e|\epsilon_{0}\quad\Longleftrightarrow\quad
    \mu\propto S^{-1/3}.
\end{equation}
This result may have interesting consequences when the source term is known
(for example the number of electrons created by an electrode per unit of 
time determinated by the Richardson-Dushman law), because it permits to 
determine the electronic mobility. Conversely, since the mobility is considered 
as a gas property, the relation (\ref{u3}) gives a first approximation of the 
value of the source term.
\newline\newline
In summary, we have presented an electrohydrodynamical model describing 
the evolution of a local electric field in a plasma. Since we make the 
assumption of motionless ions in the model, we have shown the possibility to 
derive a new equation of evolution for the electric field obeing to 
the Poisson equation. Thereby, the results presented here have to be seen as a 
continuation of Olesen's works, in the sense where we have presented and 
solved an electric version of Eqs. (\ref{b1}) and (\ref{b2}). The two basic 
equations being the equation of motion
\begin{equation}
    \pt u+u\px u=\nu\pxx u+|e|\mathcal{E},
\end{equation}
and the Poisson equation
\begin{equation}
    \px \mathcal{E}=-\frac{\vert e\vert}{\epsilon_{0}}n.
\end{equation}
Then, we have found a solution for the electric field of the form
\begin{equation}
    \label{t1}
    \mathcal{E}(x,t)=E(x,t)-E_{{{\rm{a}}}}=A(t)x+B(t),
\end{equation}
with the properties
\begin{equation}
    \label{t2}
    \lim_{t\to\infty}A(t)=-\frac{\gamma}{\mu}\quad{\rm{and}}\quad
    \lim_{t\to\infty}B(t)=0.
\end{equation}
with the meaning that the electric field behaves like an oscillating 
force in the long time limit, and with a string constant proportional to the 
square root of the source term. Furthermore, a relation may be obtained,
still in the asymptotic mode, between the terme of source $S$ and the 
electronic mobility $\mu$, and may be interpreted as a general scale law 
under some hypothesis. Notice that we have focused our attention on a 
positive term of source. The case of a negative $S$, would gives rises 
to another kind of physics which may be interesting to treat in a
forthcoming paper.

\noindent As a corollary we enforce the result discussed in ref.\cite{burg}, 
which states that the transformation 
\begin{equation}
    \label{tr1}
    u(x,t)=-2\nu\frac{1}{P(x,t)}\px P(x,t)-\kappa x,
\end{equation}
introduced in the Burgers equation (\ref{burg1}), yields to the Fokker-Planck 
equation 
\begin{equation}
    \label{fp}
    \pt P(x,t)=\kappa\px\left(xP(x,t)\right)+\nu\pxx P(x,t).
\end{equation}
And this expression may be considered as a collisional term in a Vlasov-Poisson 
model (without magnetic field) ; the electric field being determinated by 
the method presented here. Under specific assumptions, this remark 
would open an interesting way of investigation.
%%%%%%%%%%%%%%%%%%%%%%%%%%%%%
%%%%%%%%%%%%%%%%%%%%%%%%%%%%%
%%%%%%%%%%%%%%%%%%%%%%%%%%
    
\end{document}